\documentclass[12pt]{article}
\begin{document} 
\title{\bf An Alternative D$p$-brane Solution of Type IIB Supergravity }
\author{{\sc Huan-Xiong Yang \thanks{E-mail: hxyang@zimp.zju.edu.cn}, \sc Cong-Kao Wen, \sc Ben-Geng Cai}
\\
{~~}
\\
{\it Zhejiang Institute of Modern Physics, Physics Department,}
\\
{\it Zhejiang University, Hangzhou, 310027,\, P. R. China}
\\
{~~}
\\
{~~}
\\
{~~}}
\date{\today}
\maketitle

\begin{abstract}
By separating the thermal circle from the extra dimensions, we find a novel exact  D$p$-brane solution of Type IIB
supergravity, which might provide a scenario for studying the non-perturbative dynamics of QCD$_4$ from the perspective of
Type IIB supergravity.
\end{abstract}

\maketitle

\newpage

The exact solutions of 10-dimensional or 11-dimensional supergravity theories play  a crucial role in the string/M theory
phenomenological researches. Based on the conjectured gauge/gravity correspondence\cite{Maldacena}, the AdS-BH solutions of
10-dimensional Type II supergravities and 11-dimensional supergravity have been used to estimate QCD glueball mass spectra
and study other nonperturbative aspects of the low-dimensional gauge theories\cite{Witten}\cite{Koch, Csaki, Brower, Braga,
Constable}\cite{Cotrone}. The black p-brane solutions of Type $0$ and Type II supergravities have also provided proper
backgrounds for studying the inflationary physics in the framework of Mirage Cosmology\cite{Kiritsis}\cite{Pappa}.

The study of IR dynamics of the strongly coupled gauge theories based on the conjectured AdS/CFT correspondence requires the
supergravity solutions dual to non-conformal gauge theories. According to Witten's suggestion\cite{Witten}, such a dual
gravity description can be realized if we compactify the AdS metric on a thermal circle {$S^1$}. At high temperature (small
radius of the circle) the metric turns out to be an AdS black hole metric. If the thermal circle comes further from a
periodic dimension on the worldvolume of the {D$p$}-brane, we can impose anti-periodic boundary conditions on the fermionic
fields to obtain a gravity dual of the lower dimensional gauge theory so that the latter is not only conformal invariance
broken but spacetime supersymmetry broken. As a price to be paid, although Type IIB supergravity AdS$_5$-BH
solution\cite{Ohta} appeared more natural as a gravity dual to QCD$_4$, it was actually used to investigate the
nonperturbative properties of QCD$_3$\cite{Brower}. Conventionally, the IR dynamics of QCD$_4$ such as the QCD$_4$ glueball
masses was studied in the framework of AdS$_7$-BH p-brane solution of the 11-dimensional supergravity theory.

In this paper we report a novel D$p$-brane black hole solution of Type IIB supergravity. This solution describes such a
10-dimensional spacetime in which the thermal circle emerges from the compact subspace extra to the worldvolume of the
D$p$-branes. Different from the known AdS$_5$-BH solution which was usually formulated for Euclidean signature, our solution
is more natural for Minkowski signature. We expect this solution might to provide a dual description for the non-conformal
QCD$_{p+1}$-like gauge theory living on the worldvolume of the D$p$-branes.

\newpage
Our starting point is the following action of the 10-dimensional Type IIB supergravity in the Einstein frame\cite{Pol,
Petersen},
\begin{equation}\label{eq: 1}
S = -s\frac{1}{16{\pi}G_{10}}\int d^{10} x \sqrt{g} {\bigg[}R-\frac{1}{2}g^{\mu\nu}\partial_{\mu}\phi
\partial_{\nu}\phi -\frac{1}{2}\sum\limits_{p} \frac{1}{(p+2)!} e^{a_p \phi} F^2_{p+2} +\cdots{\bigg ]}~.
\end{equation}
Here \( s=-1~(+1)\) for ``mostly plus'' Minkowski (Euclidean) signature, \( G_{10}\) is the 10-dimensional Newton constant
and {\small \( a_p=(3-p)/2\)}. The dots in Eq.(\ref{eq: 1}) represent fermionic terms as well as the terms involving in NS-NS
$3$-form field strength. \( \phi \) stands for the dilaton field. The \( (p+2) \)-form field strengths \( F_{p+2}\) belong to
the RR sector. The implicit D$p$-branes form the source of RR charges for these field strengths. We shall only consider the
terms explicitly listed. For Type IIB supergravity theory, the indices \( p\) do only take odd integers from \( -1\) to \(
9\). In particular, the \( 5\)-form field strength \(F_5\) is self-dual for Minkowski signature\cite{Pol}. The action leads
to the following equations of motion for a general \( p(\ne 3)\):
\begin{equation}\label{eq: 2} \left. \begin{array}{l} R^{\mu}_{~\nu}  =
\frac{1}{2}\partial^{\mu}\phi
\partial_{\nu}\phi + \frac{1}{2} \sum\limits_p \frac{1}{(p+2)!} e^{a_p \phi}~\Big[(p+2)F^{\mu \xi_2
\cdots \xi_{p+2}} F_{\nu \xi_2 \cdots \xi_{p+2}}
-\Big(\frac{p+1}{D-2}\Big)\delta_{~\nu}^{\mu} F^2_{p+2}~\Big], ~\\
\frac{1}{\sqrt{g}} \partial_{\mu}(\sqrt{g} g^{\mu \nu}
\partial_{\nu} \phi)  =  \frac{1}{2} \sum\limits_p \frac{1}{(p+2)!} a_p e^{a_p \phi}
F_{p+2}^2,~\\
\partial_{\mu}\Big( \sqrt{g}e^{a_p \phi} F^{\mu \nu_2 \cdots \nu_{p+2}} \Big)=0,~\\
\partial_{{\bf[}\mu}F_{\mu_1 \mu_2 \cdots \mu_{p+2} {\bf]}}  =  0.
\end{array}
\right.
\end{equation}
where {\small $D=10$}. For simplicity we will consider the case with {\small $F_{p+2} \ne 0$} only for one value of {\small
$p$}. For {\small \( p=3\)}, in addition to the fact that {\small \(F_5\)} is self dual for Minkowski signature, the dilaton
is decouple from the other degrees of freedom\cite{Petersen}. The supergravity equations are then revised as:
\begin{equation}\label{eq: 3} \left. \begin{array}{l} R^{\mu}_{~\nu}  =
\frac{1}{2\cdot 5!}\Big(5F^{\mu \xi_2 \cdots \xi_5} F_{\nu \xi_2 \cdots \xi_5}
-\frac{1}{2}\delta_{~\nu}^{\mu} F^2_5~\Big), ~\\
\partial_{\mu}\Big( \sqrt{g}F^{\mu \nu_2 \cdots \nu_5} \Big)=0,~\\
(\ast F)_{\mu_1 \mu_2 \cdots \mu_5 }  = F_{\mu_1 \mu_2 \cdots \mu_5 } .
\end{array}
\right.
\end{equation}
Here the decoupled dilaton has consistently been regarded as a constant. The D$3$-brane solution of Eqs.(\ref{eq: 3}) is
expected to provide a dual description of QCD$_4$ in 4-dimensional Minkowski space-time.

We suppose that the whole 10-dimensional spacetime in Type IIB supergravity theory is a direct product {\small $K^{(1,p+2)}
\times S^d$} of a $(p+3)$-dimensional spacetime {\small $K^{(1,p+2)}$} and a {\small $(d-2)$}-dimensional sphere {\small
$S^{d-2}~(p+d=D-1)$}. Besides the {\small $(p+1)$}-dimensional sub-spacetime {\small $M^{(1,p)}$} occupied by {\small
D$p$}-branes, there are an extra dimension and a thermal circle {\small $S^1$} in {\small $K^{(1,p+2)}$}. We label the
spacetime by the coordinates {\small $z^{M}=(\theta, r, t, x^{i}, y^{\alpha})$}, where {\small $i=1, 2, \cdots,p$} and
{\small \( \alpha=1, 2, \cdots, d-2 \)}. The coordinate of extra dimension coincides with the squared radius {\small \( r =
\sqrt{\sum_{\alpha}(y^{\alpha})^2} \)} of the transverse sphere {\small $S^{d-2}$}. The metric respecting to the symmetries
of above 10-dimensional spacetime reads:
\begin{equation}\label{eq: 4}
ds^2 = B^2(r) d\theta^2 + C^2(r) \Big[sdt^2 + \sum^{p}_{i=1}(dx^i)^2 \Big]~+F^2(r) dr^2 +G^2(r) r^2 \big(d\Omega_{d-2}\big)^2
~.
\end{equation}
This is a diagonal metric where all the components are assumed to depend only on the transverse distance {\small $r$}. The
SO(1,~p) invariance of the metric ansatz (\ref{eq: 4}) might imply that the probable solutions of Eqs.(\ref{eq: 2}-\ref{eq:
3}) considered here are generally independent of those in \cite{Ohta}.  In terms of Eq.(\ref{eq: 4}),
\begin{equation}\label{eq: 5}
\left.
\begin{array}{lcl}
R^{\theta}_{~\theta} & = & - F^{-2} [(\ln B)^{\prime \prime} + (\ln
B)^{\prime} (\ln fr^{d-2})^{\prime}] \\
R^{r}_{~r} & = & - F^{-2} \{(\ln Ffr^{d-2})^{\prime \prime} - (\ln
F)^{\prime} (\ln Ffr^{d-2})^{\prime} +[(\ln B)^{\prime}]^2 \\
& & ~~~~~+ (p+1)[(\ln C)^{\prime}]^2 + (d-2)[(\ln Gr)^{\prime}]^2\} \\
R^{t}_{~t} & = & R^{\underline{i}}_{~\underline{i}} = - F^{-2} [(\ln C)^{\prime \prime} + (\ln
C)^{\prime} (\ln fr^{d-2})^{\prime}] \\
R^{\underline{\alpha}}_{~\underline{\alpha}} & = & - F^{-2} [(\ln Gr)^{\prime \prime} + (\ln Gr)^{\prime} (\ln
fr^{d-2})^{\prime} -(d-3)\frac{F^2}{(Gr)^{2}} ]~.
\end{array}
\right.
\end{equation}
where $f=BF^{-1}C^{p+1}G^{d-2}$ and $(\ln B)^{\prime}$, say, denotes the derivative of function $(\ln B)$ with respect to the
coordinate $r$.

For a general {\small $p~(\ne 3)$}, the {\small $(p+1)$}-form RR gauge potential could naturally couple to the worldvolume of
the D{\small $p$}-branes, in which the resulting {\small $(p+2)$}-form RR field strength is termed as electric. We assume the
following electric ansatz for the RR field strength {\small $F_{p+2}$}:
\begin{equation}\label{eq: 6}
F_{t i_1 i_2 \cdots i_p r}(r)= \delta^{1 2~ \cdots ~p}_{i_1 i_2 \cdots i_p}  \Psi(r).
\end{equation}
The Banchi identities [the third set of equations in Eq.(\ref{eq: 1})] are automatically satisfied by this ansatz. The field
equation of RR field strength leads to:
\begin{equation}\label{eq: 7}
\Psi(r)=e^{-a_p \phi}\frac{C^{p+1}FQ}{B(Gr)^{d-2}},
\end{equation}
with $Q$ the constant of integration\cite{Petersen}. Consequently,
\begin{equation}\label{eq: 8}
\left. \begin{array}{rcl} \frac{1}{(p+2)!}F^2_{p+2} & = &se^{-2a_p \phi}\frac{Q^2}{B^2 (Gr)^{2d-4}},\\
\frac{1}{(p+1)!}F^{\mu \xi_2 \cdots \xi_{p+2}} F_{\nu \xi_2 \cdots \xi_{p+2}} & = & s \delta^{\mu}_{~\nu}e^{-2a_p
\phi}\frac{Q^2}{B^2 (Gr)^{2d-4}}.
\end{array}
\right.
\end{equation}
with $\mu, \nu \in \{t, r, 1,2,\cdots, p\}$ in the second equation. We can recast the Einstein equations in Eqs.(\ref{eq: 2})
as:
\begin{equation}\label{eq: 10}
\left. \begin{array}{lll}
R^{r}_{~r} & = & \frac{{\phi^{\prime}}^2}{2F^2} + R^{t}_{~t}\\
R^{t}_{~t} & = & R^{\underline{i}}_{~\underline{i}} = - (\frac{d-2}{p+1})R^{\theta}_{~\theta} \\
R^{\theta}_{~\theta} & = & R^{\underline{\alpha}}_{~\underline{\alpha}}\\
R^{\underline{\alpha}}_{~\underline{\alpha}} & = & - se^{-2a_p \phi} \frac{(p+1)Q^2}{2(D-2)B^2 (Gr)^{2d-4}}.
\end{array} \right.
\end{equation}
where $\underline{i}=1,2, \cdots, p$ and $\underline{\alpha}$ stands for an arbitrary coordinate on $S^{d-2}$. They yield the
following differential equations for determining the metric components in Ansatz (\ref{eq: 4}):
\begin{equation}\label{eq: 11}
(\ln B^{d-2}C^{p+1})^{\prime \prime} + (\ln B^{d-2}C^{p+1})^{\prime}\Big[(\ln f)^{\prime} +(d-2)/r\Big]=0.
\end{equation}
\begin{equation}\label{eq: 12}
\left. \begin{array}{ll} (\ln G^{d-2}C^{p+1})^{\prime \prime} &+ (\ln G^{d-2}C^{p+1})^{\prime} (\ln fr^{d-2})^{\prime}
+\frac{(d-2)}{r}(\ln f)^{\prime} \\
& +\frac{(d-2)(d-3)}{r^2} [1- ({F}/{G})^2]  =0.
\end{array}
\right.
\end{equation}
\begin{equation}\label{eq: 13}
\left. \begin{array}{ll} (\ln Ffr^{d-2})^{\prime \prime} & +  (\ln F)^{\prime} (\ln Ffr^{d-2})^{\prime} +{(\ln B)^{\prime}}^2
+(p+1){(\ln C)^{\prime}}^2 \\ & +(d-2){(\ln Gr)^{\prime}}^2 -(\ln C)^{\prime \prime} -(\ln C)^{\prime}(\ln fr^{d-2})^{\prime}
+\frac{1}{2} (\phi^{\prime})^2  =0.
\end{array}
\right.
\end{equation}
\begin{equation}\label{eq: 14}
(\ln Gr)^{\prime \prime} + (\ln Gr)^{\prime} (\ln fr^{d-2})^{\prime} -(d-3)[F/(Gr)]^2 -se^{-a_p \phi}\frac{(p+1)F^2
Q^2}{2(D-2)B^2 (Gr)^{2d-4}}  =0.
\end{equation}
Besides,the field equation of dilaton is also recast as:
\begin{equation}
\label{eq: 15} \phi^{\prime \prime} + \frac{[f+(d-3)]}{fr} \phi^{\prime} -sa_p e^{-a_p \phi} \frac{Q^2}{2f^2 (Gr)^{2d-4}}=0.
\end{equation}

We now intend to find some exact solutions to these equations. Let
\begin{equation}\label{eq: 16}
\ln B^{d-2}C^{p+1} =c_B \ln f,
\end{equation}
with $c_B$ a constant, we see from Eq.(\ref{eq: 11}),
\begin{equation}\label{eq: 17}
f^{\prime \prime} + \frac{(d-2)}{r} f^{\prime}=0.
\end{equation}
By imposing the asymptotic condition $\lim\limits_{r \rightarrow \infty} f(r)=1$, this equation enable us to determine the
function $f(r)$:
\begin{equation}\label{eq: 18}
f=1-(r_0/r)^{d-3}
\end{equation}

Eq.(\ref{eq: 12}) can be reduced to
\begin{equation}\label{eq: 19}
(\ln f)^{\prime} + \frac{(d-3)}{r}(1-f^{2c_F})=0
\end{equation}
if we assume $G^{d-2}C^{p+1}=1$ and $F=Gf^{c_F}$. The consistency between Eq.(\ref{eq: 19}) and its solution (\ref{eq: 18})
indicates $c_F =-1/2$ and furthermore $c_B=(d-2)/2$. These data enable us to express all the metric components in terms of a
unique function $G(r)$:
\begin{equation}\label{eq: 20}
F=G/\sqrt{f},~~B=G\sqrt{f},~~C=G^{-(\frac{d-2}{p+1})}.
\end{equation}
By combining Eq.(\ref{eq: 20}) and Eqs.(\ref{eq: 13}-\ref{eq: 15}) we find that dilaton is also related to {\small $G(r)$}:
{\small $ \phi = a_p (\frac{D-2}{p+1})\ln G. $} Let
\begin{equation}\label{eq: 21}
G = H_p^{\frac{p+1}{\Delta}}~
\end{equation}
with {\small $\Delta = (d-2)(p+1)+ \frac{1}{2}a^2_p (D-2)$}. The metric turns out to be determined by the equations,
\begin{equation}\label{eq: 22}
\left. \begin{array}{ll} & H_p^{\prime \prime} + \frac{f+(d-3)}{fr}H_p^{\prime}=0\\
& {H_p^{\prime}}^2 +s \frac{\Delta Q^2}{2(D-2)f^2 r^{2(d-2)}}=0
\end{array}
\right.
\end{equation}
The exact solution to Eqs.(\ref{eq: 22}) reads:
\begin{equation}\label{eq: 23}
H_p(r)= \left \{
\begin{array}{ll}
1+\frac{h}{(d-3)r^{d-3}_0}\ln[1-(r_0/r)^{d-3}],~~&\textrm{if $r^0 \ne 0$}\\
1-\frac{h}{(d-3)r^{d-3}} ,~~&\textrm{if $r_0=0$}
\end{array}
\right.
\end{equation}
where the integration constant {\small $h$} is subject to an algebraic equation {\small $h^2 + s\frac{\Delta Q^2}{2(D-2)}=0$
}, which is real when we take {\small $s=-1$}.

Now we consider the case {\small $p=3$} with Minkowski signature. The above conclusion is almost applicable if we let {\small
$p=3$} in the listed equations except that the dilaton is decouple from the metric tensor and the RR field strength {\small
$F_5$} must be self-dual. Instead of the simple assumption (\ref{eq: 6}) for {\small $F_{p+2} ~(p\ne 3)$}, the self-duality
characteristic of {\small $F_5$} enforce us to suppose
\begin{equation}\label{eq: 241}
F_5 = \frac{FC^4 Q}{B(Gr)^4}dt\wedge dx^1 \wedge dx^2 \wedge dx^3 \wedge dr + \sqrt{\gamma_4}Q d\theta \wedge d\alpha^1
\wedge d\alpha^2 \wedge d\alpha^3 \wedge d\alpha^4
\end{equation}
where {\small $\alpha^{i}~(i=1,\cdots,4)$} are the local coordinates on $S^4$ and {\small $\gamma_4$} stands for the
determinant of the corresponding metric tensor. Eq.(\ref{eq: 241}) indicates {\small $\frac{1}{5!}F_5^2 =0$} while {\small
$\frac{1}{4!}F^{\alpha \xi_2 \cdots \xi_5}F_{\beta \xi_2 \cdots \xi_5} =\delta^{\alpha}_{~\beta}\frac{Q^2}{B^2(Gr)^8}$} for
{\small $\alpha, \beta \in \theta, \alpha_1, \cdots, \alpha_4$}. However, such a difference does not change the mathematical
structure of the solution (\ref{eq: 23}). A remarkable alteration is that the integration constant {\small $h$} is shifted to
{\small $h =\pm \sqrt2 Q$}.

In conclusion, we have found a novel Dp-brane black hole solution to the bosonic action (\ref{eq: 1}) of Type IIB
supergravity. It is spherically symmetric with all the fields are only dependent upon the radial coordinate of the extra
dimensions. The solution reads
$$
\left. \begin{array}{lll} ds^2_{10} & = & H_p^{\frac{2(p+1)}{\Delta}}f(d\theta)^2 +H_p^{\frac{2(p+1)}{\Delta}}f^{-1}(dr)^2 +
H_p^{-\frac{2(d-2)}{\Delta}}[s(dt)^2 + \sum_{i=1}^{p} (dx_i)^2]\\
&  & + H_p^{\frac{2(p+1)}{\Delta}}r^2 (d\Omega_{d-2})^2,
\end{array}
\right.
$$
\begin{equation}\label{eq: 24}
\phi  = (\frac{D-2}{\Delta})a_p \ln H_p ,~~~~F_{p+2}=\frac{Q}{fH^2 r^{d-2}} dt\wedge dx^1 \wedge \cdots \wedge dx^p \wedge
dr~.
\end{equation}
for a general {\small $p(\ne 3)$} and
$$
\left. \begin{array}{lll} ds^2_{10} & = & \sqrt{H_3} [1-(r_0/r)^3](d\theta)^2 +\frac{\sqrt{H_3 }}{[1-(r_0/r)^3]}(dr)^2 +
\frac{1}{\sqrt {H_3}} [-(dt)^2 + \sum_{i=1}^{3} (dx_i)^2]\\
&  & + \sqrt{H_3} r^2 (d\Omega_4)^2,
\end{array}
\right.
$$
\begin{equation}\label{eq: 25}
\phi  = 0,~~F_5 = \Big[ \frac{Q}{H^2_3 r(r^3-r^3_0)} \Big] dt\wedge dx^1 \wedge dx^2 \wedge dx^3 \wedge dr + \sqrt{\gamma_4}Q
d\theta \wedge d\alpha^1 \wedge d\alpha^2 \wedge d\alpha^3 \wedge d\alpha^4.
\end{equation}
for {\small $p=3$}. If {\small $r_0=0$}, this background metric in Eqs.(\ref{eq: 24}-\ref{eq: 25}) approaches to that of the
spacetime {\small $AdS_{p+2} \times S^{d-2} \times S^{1}$} in the limit of {\small $r \rightarrow 0$}. While for a general
{\small $r_0 (\ne 0)$}, the metric defines a 10-dimensional spacetime that has an extra compact dimension {\small $S^1$}
transverse to the worldvolume of the D{\small $p$}-branes. This compact dimension might be a reasonable candidate for the
thermal circle if Type IIB supergravity could provide a dual description for the nonperturbative aspects of the 4-dimensional
Yang-Mills gauge theories such as {\small QCD$_4$}.

\subsection*{Acknowledgments}
We would like to thank M.X.Luo, N.Ohta, Q.P.Su and W.S.Xu for valuable discussions. The work is supported in part, by
CNSF-10375052, the Startup Foundation of the Zhejiang Education Bureau and CNSF-90303003. H.X. Yang has also benefitted from
the support of Pao's Foundation.

\end{document}